# Microwaves

## *Microwave-Assisted Hydrothermal Synthesis of Nanoparticles*

*Rainer Schmidt, Jesús Prado Gonjal, and Emilio Morán*

### CONTENTS

### INTRODUCTION

The general topic of *energy* is regarded to be one of the most important areas for any research and development activities at the start of the twenty-first century. Whereas the worldwide energy consumption and the concomitant $CO_2$ emissions are increasing continuously, the resulting climate changes have been identified to be a major threat upon reaching irreversible levels. Therefore, large research efforts are already based on or will be targeted on the development of clean fuels to replace fossil fuels, the improvement of the existing resources and the discovery of new sustainable energy resources, the establishment of new efficient ways to store sustainable energy, and the replacement of the existing high-energy-consuming technologies with the more energy-efficient ones.

The latter concept has resulted in considerable efforts recently to renovate the conventional synthetic chemical industry by introducing new *green chemistry* concepts, where the energy requirements for chemical processing or synthesis would be drastically reduced. In this con- text, a vast amount of novel chemical synthesis techniques have been developed during the last decades in an effort to reduce energy consumption by reduced chemical reaction temperatures and reaction times. Such innovative techniques include combustion, sol–gel and coprecipitation methods, sonochemistry, hydrothermal synthesis, and microwave-assisted techniques.

The considerable reduction in reaction temperature and time involved with the recent innovative synthesis techniques has the positive *side effect* that the particle growth during reaction and the resulting particle size of the synthesized products are usually reduced efficiently. This not only is interesting for potential new applications of nanosized materials due to novel functionalities but also facilitates studying fundamental aspects of the physics of condensed matter at the nanoscale. Significant qualitative drawbacks in materials synthesized by such novel ways in terms of crystal quality and physical properties are usually not encountered. In some cases, even quantitative improvements, novel crystal arrangements, and interesting particle shapes can be achieved.[1–8]

The concept of combining several of the recently developed nonconventional and innovative synthesis techniques is an interesting concept to even further reduce the energy requirements and particle sizes as compared to one technique alone. Microwave-assisted combustion and sol–gel synthesis are now employed frequently, as well as the combination of hydrothermal and microwave-assisted synthesis, which is the topic of this review.

Microwave-assisted and hydrothermal synthesis as innovative synthesis techniques on their own can be characterized separately as follows:

1. Hydrothermal synthesis refers to materials synthesis by chemical reactions in aqueous media at

intermediate temperatures ($T \approx 200°C$) in a pressure-sealed reaction vessel. Heating the appropriate precursor solutions in the vessel results in high pressure. This increases the dispersion of the system components in a compressed liquid environment, and lower temperatures and shorter reaction times are sufficient to complete the reaction as compared to conventional synthesis. To even further improve the dispersion of the system in solution, magnetic stirring is commonly employed during typical hydrothermal synthesis reactions.

2. Microwave synthesis refers to the utilization of microwave radiation as a heat source for materials synthesis. In contrast to conventional heat transfer by heat convection, microwaves directly couple to the material via several possible mechanisms; thus, the method is ideally suited to directly transfer heat to the reactants. The main microwave heating mechanisms are microwaves directly coupling to electric dipoles and dielectric and magnetic loss heating. Due to the direct and intense heat transfer, the reaction completes quickly, far away from thermal equilibrium. Since microwaves couple well to electric dipoles, the high polarity of $H_2O$ in an aqueous solution for hydrothermal synthesis is ideally suited for microwave heat exposure.

Due to strongly reduced reaction time in combined microwave-assisted hydrothermal synthesis (Figure 1), this technique can be regarded a *fast chemistry* method. Furthermore, taking into account the low-energy requirements, the technique is consistent with most of the *green chemistry* principles: use of nonhazardous reactants and solvents, high efficiency in terms of energy consumption versus yield, and easy monitoring to prevent pollution.[9,10]

This work reviews the basic principles of the combination of hydrothermal and microwave-assisted syntheses for energy-efficient *green chemistry* synthesis of nanoparticles.

## BASIC PRINCIPLES OF HYDROTHERMAL SYNTHESIS

The method of hydrothermal synthesis on its own is a well-established *wet-chemical*, moderate-pressure method of synthesis, which consists in an aqueous solution that is heated to $T \approx 200°C$ in a sealed reaction vessel (autoclave), and concomitantly the pressure $p$ increases due to the fixed volume of the vessel. This results in an increase of the dispersion of the system components in a compressed liquid environment, and the reaction completes more quickly and at lower temperature. It should be noted that other solvents than water can be used in the so-called solvothermal synthesis method, which has become increasingly popular in recent years. For hydrothermal synthesis with water solvents, it is important to consider *subcritical* and *supercritical* synthesis conditions, where the temperature is below or above the critical temperature of water ($T_{cr} = 374°C$), respectively. By increasing the temperature above $T_{cr}$, the pressure in the synthesis autoclave can raise in an ungoverned fashion, which makes the reaction process difficult to control (Figure 2).[11,12]

The main difference between hydrothermal and solid-state reaction lies in the reactivity that is reflected in their different reaction mechanisms. Solid-state reactions depend on the diffusion of the raw materials at their interfaces, whereas in hydrothermal or solvothermal reactions, the reactant ions and/or molecules react in a compressed liquid environment where dispersion of the reactants is far superior.

Difference in the reaction ambient can lead to different microstructures of the final products, even if the same reactants are used.[11] Therefore, hydrothermal synthesis requires the careful adjustment of the critical parameters such as the pH of the solution, the reaction temperature, and pressure. Although the hydrothermal method is very versatile, one of the main drawbacks is the slow kinetics of crystallization. In order to increase the kinetics of crystallization, one may use microwaves for heating the solution to accelerate crystallization kinetics. Such increased reaction rates are debited to the efficient and the direct heat transfer offered by microwaves due to excellent coupling of the microwave to the water molecules in the solution.[14] Details of such coupling are discussed in the section "Interactions of Microwaves and Matter."

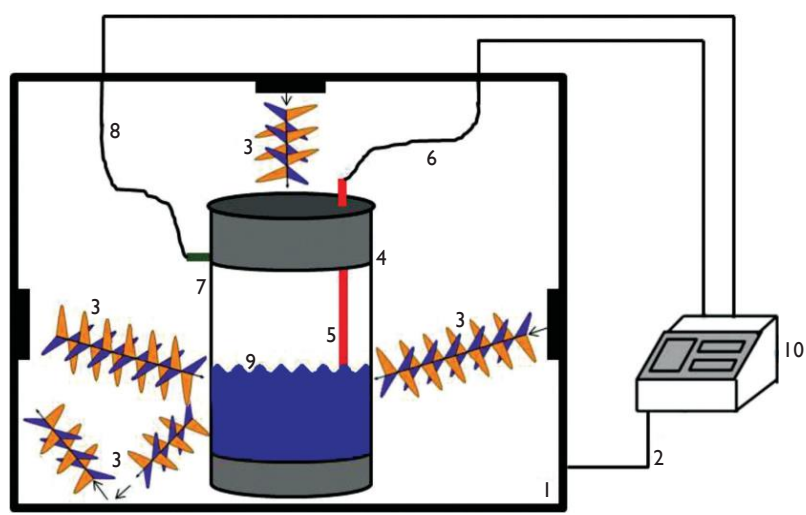

**FIGURE 1** Schematic drawing of the principles of a microwave-assisted hydrothermal synthesis process: (1) microwave oven, (2) microwave power control line, (3) idealized microwave radiation, (4) pressure sealed reaction vessel, (5) temperature sensor, (6) temperature control line, (7) pressure gauge, (8) pressure control line, (9) reactants in solution, and (10) control line.

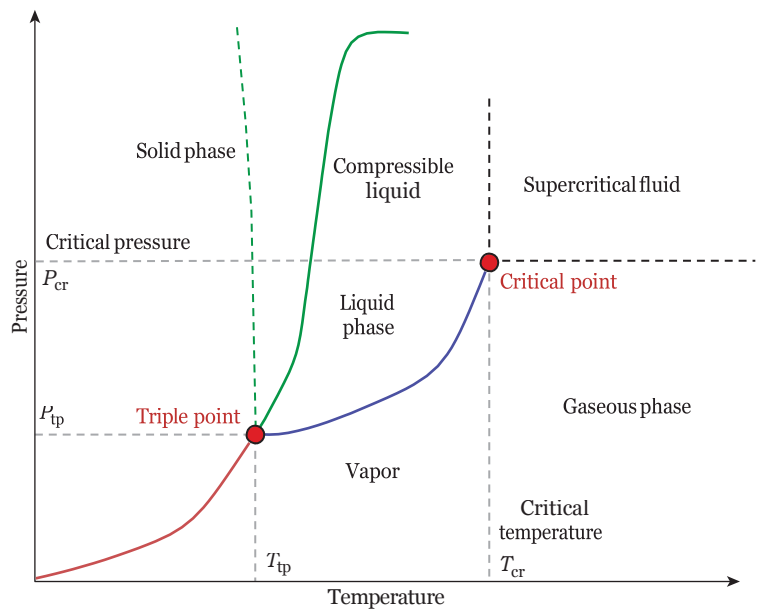

**FIGURE 2** Phase diagram of water. (From Critical point (thermodynamics), http://en.wikipedia.org/wiki/Critical_point_(thermodynamics), 2013.)

## APPARATUS FOR MICROWAVE-ASSISTED HYDROTHERMAL SYNTHESIS

In the early 1990s, Dr. Sridhar Komarneni at the University of Pennsylvania (United States) launched his pioneering work by studying and comparing the differences between hydrothermal syntheses performed with conventional means of heating and hydrothermal syntheses performed by heating special reaction vessel autoclaves with microwaves.[15,16] The main characteristic microwave parameters are power and time, which need to be adjusted carefully for the desired reaction temperature and time in the hydrothermal reaction vessels as mentioned previously. The reaction temperature and pressure in the autoclaves resulting from microwave irradiation are difficult to predict from the microwave power and exposure time alone. Therefore, advanced synthesis technology is required to closely control pressure and temperature in the subcritical region of water in situ during the reaction. A typical commercial furnace used for microwave hydro-/solvothermal processing is shown in Figure 3. There are several companies that distribute suitable furnaces for microwave-assisted hydrothermal synthesis: Milestone, CEM, Biotage, MARS, and Anton Paar.

A typical microwave-assisted hydrothermal synthesis apparatus is usually equipped with advanced reaction sensors, including direct contactless temperature and pressure control in the reaction and a control vessel (see right side of

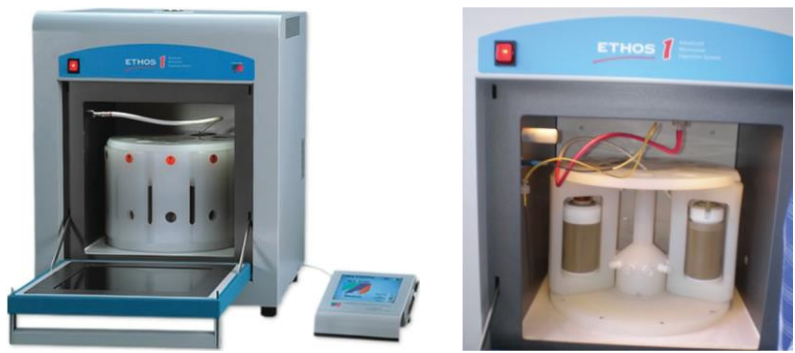

**FIGURE 3** Commercial Milestone Ethos One furnace for microwave-assisted hydrothermal synthesis.

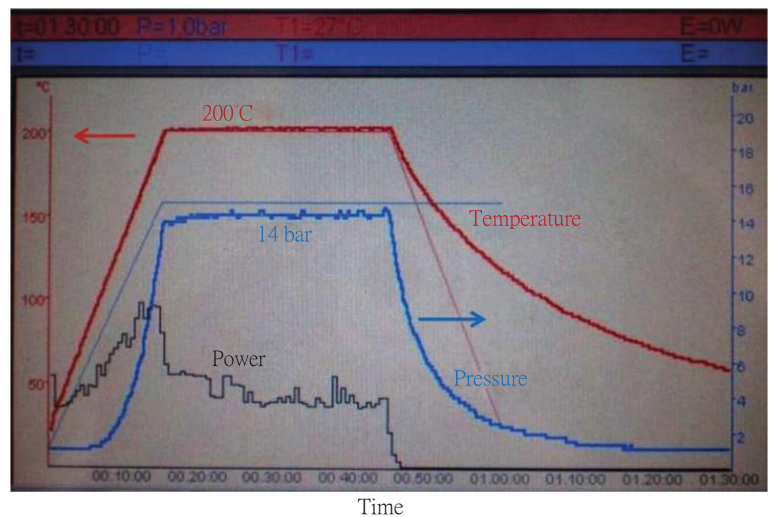

**FIGURE 4** Typical profile of the reaction parameters temperature, pressure, and microwave power during a microwave-assisted hydrothermal synthesis process during 30 min at 200°C and 14 bars.

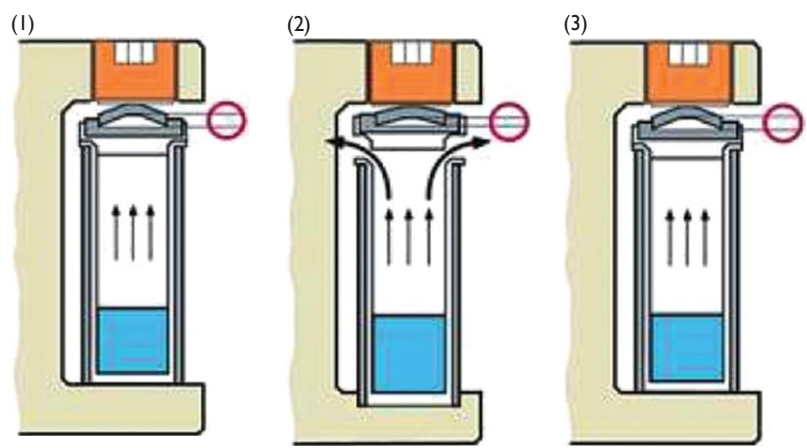

**FIGURE 5** *Vent-and-reseal* Milestone technology: The *dome-shaped* spring is temporarily flattened in case of overpressure (1), allowing the cap to lift up slightly and release the pressure (2). Once the pressure has been released, the cap closes again (3), and the microwave program can continue.

Figure 3).[17] This allows monitoring the reaction conditions in situ as demonstrated in Figure 4, depicting the typical profile of the critical parameters for a representative 30 min hydrothermal synthesis process at 200°C and 14 bars.

Autoclaves are made out of Teflon, in which case the temperature and pressure are limited to 250°C and 100 bars or out of quartz (≥250°C, 45 bars). The Milestone system releases potential excess pressure from the vessel using the *vent-and-reseal* vessels technology (U.S. Patent 5,270,010).[18] This technology demonstrated in Figure 5 ensures that there is no stress to the autoclave vessels, which, in the worst case, can lead to a vessel bursting.

In contrast to domestic ovens, dedicated microwave-assisted hydrothermal reactors feature a built-in magnetic stirrer, which is of great importance since the temperature distribution within the reaction mixture may not be uniform without proper agitation and the measured temperature will be dependent on the position of the temperature sensor, that is, stirring helps the system to quickly disperse any localized superheating, that is, the formation of *hot spots* can be avoided.[19] Microwaves couple directly with the reactants and solvent mixture, leading to a rapid rise in temperature (Figure 6), and the occurrence of hot spots is in fact quite common in microwave synthesis.

## ENERGY TRANSFER BY COUPLING OF MICROWAVES TO THE SOLVENT

When microwave irradiation is combined with the hydrothermal method, an important factor is the choice of solvent.[21] The ability of a solvent to couple to the microwave determines the efficiency of microwave energy absorption by the reaction mixture and the outcome of the reaction. More efficient coupling leads to faster increase of the temperature. It is mentioned later (in the section "Interactions of Microwaves and Matter") that several mechanisms exist for electromagnetic microwaves to couple and directly transfer energy to the reactants. However, for hydrothermal synthesis of a nonmagnetic system, only two dielectric parameters of the solvent are the main factors to define the ability of the reaction mixture to absorb microwave energy[20,22]:

1. Dielectric constant ($\varepsilon'$)
2. Dielectric losses ($\varepsilon''$)

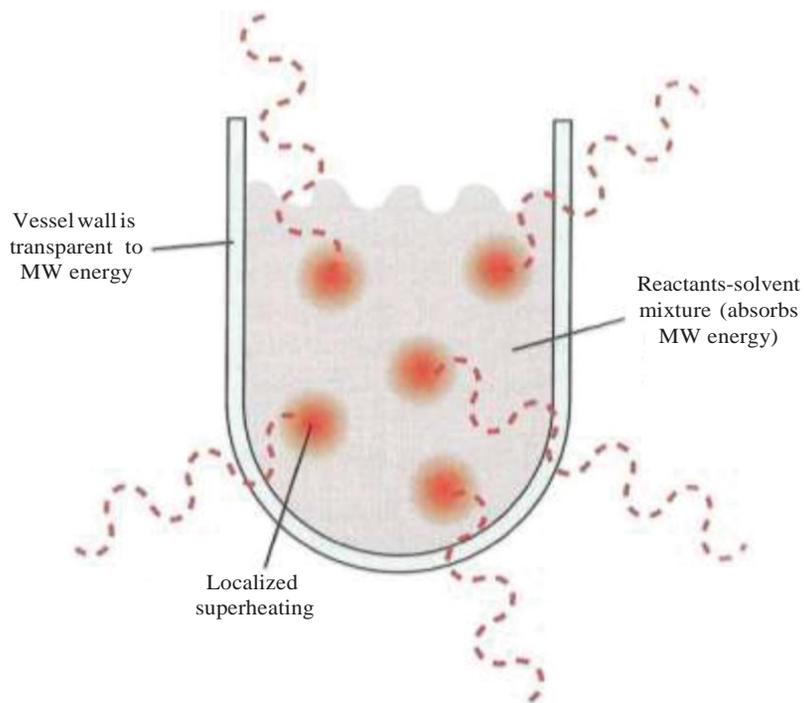

**FIGURE 6** Schematic of sample heating by microwave-assisted hydrothermal method. (From Hayes, B.L., *Microwave Synthesis: Chemistry at the Speed of Light*, CEM Pub., Matthews, NC, p. 295, 2002.)

$\varepsilon'$ should be generally as high as possible to increase the efficiency of the coupling mechanism between microwave radiation and the dipoles in the solvent, whereas the optimum value for $\varepsilon''$ is less straightforward. Polar solvents with losses $\varepsilon'' < 10^{-2}$ are quite difficult to be heated up, while for those with $\varepsilon'' > 5$ the surface is heated rather than the interior. In liquid hydrothermal systems, the predominant heating of the solvent surface is not critical though, because it can be compensated by homogenizing the temperature of the reaction mixture by magnetic stirring. Contrarily, in solid-state microwave synthesis, only materials with intermediate $\varepsilon''$ values of $10^{-2} < \varepsilon'' < 5$ are good candidates for uniform microwave heating across the entire bulk solid.[23,24]

The most commonly used microwave frequency in microwave-assisted synthesis is 2.45 GHz, where all dielectric parameters strongly depend on the temperature. As the temperature of the solvent increases, clear changes in the dielectric parameters can be observed, and hence the coupling efficiency between microwave and solvent changes. Graphical representations of the temperature dependence of $\varepsilon'$ and $\varepsilon''$ in several standard solvents are shown in Figures 7 and 8.[20]

Solvents can easily be categorized into three different groups, examining the values of the dielectric parameters. High-microwave-absorbing solvents exhibit dielectric losses higher than 14, which leads to increased surface heating, and magnetic stirring is necessary. Medium absorbers have dielectric losses between 1 and 14, and low absorbing molecules have less than 1 that leads to poor heating (Table 1).[20]

Water has the highest dielectric constant (80.4) of the list of solvents, and dielectric losses are medium ($\varepsilon'' = 9.889$, at room temperature and 2.45 GHz). Thus, it should be classified as a medium absorber and is ideally suited for microwave heating.

## APPLICATIONS OF MICROWAVE-ASSISTED HYDROTHERMAL SYNTHESIS

Microwave-assisted hydrothermal synthesis has successfully been applied for a vast range of materials, since the method has basically no restrictions for any type of material. In the literature, reports can be found on the synthesis of different materials ranging from binary metallic oxides, oxyhydroxides, and ternary oxides to more complex materials and structures such as zeolites or other mesoporous materials. It is an effective method for producing nanoparticles and controlling the particle size through the nucleation and growth kinetics. A careful adjustment of the pH value of the initial solution, the intensity of the magnetic stirring of the solution during reaction, the reaction temperature, pressure, and time usually leads to a certain range of particle sizes available in the final product depending on the specific system, solvents, and the starting materials used. The sensitivity of the reaction ambient also allows good control on the shape of the resulting products such as spherical, cubic, and fiber- or platelike structures as well as the formation of nanochains and core–shell structures.

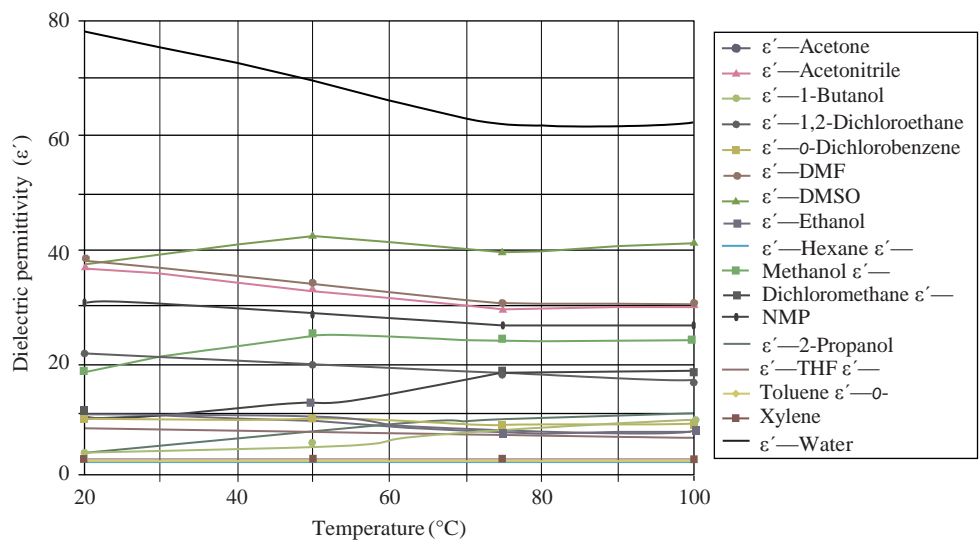

**FIGURE 7** Dielectric constant ε′ versus temperature. (From Hayes, B.L., *Microwave Synthesis: Chemistry at the Speed of Light*, CEM Pub., Matthews, NC, p. 295, 2002.)

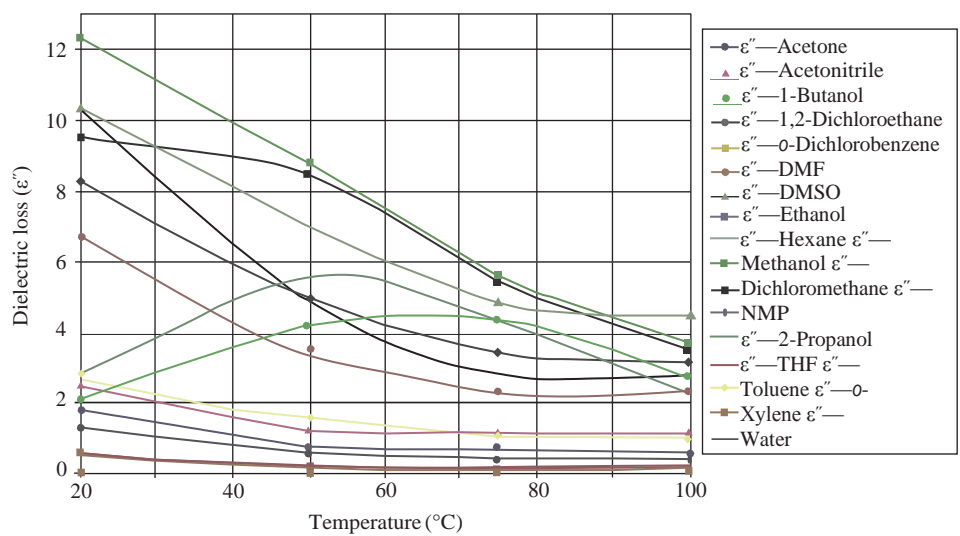

**FIGURE 8** Dielectric losses ε″ versus temperature. (From Hayes, B.L., *Microwave Synthesis: Chemistry at the Speed of Light*, CEM Pub., Matthews, NC, p. 295, 2002.)

Successful prediction tools for the particle sizes and shapes based on the reaction conditions have not been reported in the literature so far, and the parameter adjustments may well be a trial-and-error process. Some selected phases are listed in the following, which have been synthesized by microwave-assisted hydrothermal synthesis:

1. Binary metallic oxides: ZnO, CuO, PdO, CoO, MnO, $TiO_2$, $CeO_2$, SnO2, $HfO_2$, $ZrO_2$, $Nd_2O_3$, $In_2O_3$, $Tl_2O_3$, $Fe_2O_3$, $Fe_3O_4$, and $Mn_3O_4$.[3,15,25–29]
2. Oxyhydroxides: γ-AlOOH nanoparticles, α-FeOOH hollow spheres, or β-FeOOH architectures have been synthesized, where close control of temperature and pressure is particularly important.[30–34]
3. Complex oxides such as perovskites and spinels: $KNbO_3$, $ATiO_3$ (A = Ba, Pb), $NaTaO_3$, $BiFeO_3$,[15,35–38] and $ZnM_2O_4$ (M = Al, Ga) and $AFe_2O_4$ (A = Zn, Ni, Mn, Co).[39,40]
4. Nanoporous materials such as zeolites: G. A. Tompsett et al.[41] mentioned aluminosilicates and aluminophosphates structures such as MFI, sodalite, NaY (FAU), LSX (FAU), analcime, Na-P1 (Gis), Ti-ZSM-5 (MFI), AlPO4, Toad, cloverite, MnAPO-5VAPSO-44, MnAPO-5, MCM-41, SBA-15, UDF-1, and PSU-1.

This list is rather incomplete, because microwave-assisted hydrothermal synthesis has been applied in all sectors of synthetic chemistry. As mentioned earlier, the method also offers a great variety of different particle sizes in the nanometer to

**TABLE 1**

**Dielectric Constant ε′ and Dielectric Losses ε″ for Solvents Commonly Used in Microwave-Assisted Hydrothermal Synthesis, Measured at Room Temperature and 2.45 GHz**

| Solvent (Boiling Point [ºC]) | Dielectric Permittivity ($\varepsilon'$) | Dielectric Loss ($\varepsilon''$) |
|---|---|---|
| Water (100) | 80.4 | 9.889 |
| Formic acid (100) | 58.5 | 42.237 |
| DMSO (189) | 45.0 | 37.125 |
| DMF (153) | 37.7 | 6.070 |
| Acetonitrile (82) | 37.5 | 2.325 |
| Ethylene glycol (197) | 37.0 | 49.950 |
| Nitromethane (101) | 36.0 | 2.304 |
| Nitrobenzene (202) | 34.8 | 20.497 |
| Methanol (65) | 32.6 | 21.483 |
| NMP (215) | 32.2 | 8.855 |
| Ethanol (78) | 24.3 | 22.866 |
| Acetone (56) | 20.7 | 1.118 |
| 1-Propanol (97) | 20.1 | 15.216 |
| MEK (80) | 18.5 | 1.462 |
| 2-Propanol (82) | 18.3 | 14.622 |
| 1-Butanol (118) | 17.1 | 9.764 |
| 2-Methoxyethanol (124) | 16.9 | 6.929 |
| 2-Butanol (100) | 15.8 | 7.063 |
| Isobutanol (108) | 15.8 | 8.248 |
| 1,2-Dichloroethane (83) | 10.4 | 1.321 |
| o-Dichlorobenzene (180) | 9.9 | 2.772 |
| Dichloromethane (40) | 9.1 | 0.382 |
| THF (66) | 7.4 | 0.348 |
| Acetic acid (113) | 6.2 | 1.079 |
| Ethyl acetate (77) | 6.0 | 0.354 |
| Chloroform (61) | 4.8 | 0.437 |
| Chlorobenzene (132) | 2.6 | 0.263 |
| o-Xylene (144) | 2.6 | 0.047 |
| Toluene (111) | 2.4 | 0.096 |
| Hexane (69) | 1.9 | 0.038 |

*Source:* Hayes, B.L., *Microwave Synthesis: Chemistry at the Speed of Light*, CEM Pub., Matthews, NC, p. 295, 2002.

the micrometer range and various shapes of the final products. Several good review articles exist to account for this vast range of different materials with different particle size and shapes that can be obtained.[1,2,41–45]

## INTERACTIONS OF MICROWAVES AND MATTER

From a fundamental point of view, microwaves are electromagnetic waves that consist of a kind of pure energy radiated in the form of a wave propagating at the speed of light $c$. An electromagnetic wave is formed by both alternating electric and magnetic fields that oscillate at right angles to each other in phase at the same frequency in the direction of propagation.[46] The microwave propagates at lower speed within condensed matter than in air or vacuum, where $c$ is lower. This is because the speed of light $c$ within a material is connected to several material-specific parameters via the well-known relation $c = (\varepsilon_0 \varepsilon' \mu_0 \mu')^{-1/2}$, where $\varepsilon_0$ and $\varepsilon'$ correspond to the vacuum and relative dielectric permittivity and $\mu_0$ and $\mu'$ are the vacuum and relative magnetic permeability, respectively.

At a first glance, the efficient energy transfer between microwaves and the reaction mixture in a microwave-assisted hydrothermal synthesis process and the resulting heating effects are quite surprising, because the microwave radiation wavelength ranges in the low-energy zone of the electromagnetic spectrum. As shown in Table 2, a typical microwave frequency of 2.45 GHz is equivalent to $10^{-5}$ eV, which is far below the energy corresponding to a weak bond energy such as the hydrogen molecule bond of 0.21 eV.[47]

A full understanding of the energy transfer between microwave and matter is rather complicated though, because the microwave heating effect can rely on up to five different mechanisms. Such mechanisms are as follows:

**TABLE 2**
**Energy of Common Radiation**

|  | Typical Frequency (Mz) | Quantum Energy (eV) |
|---|---|---|
| γ-rays | 3.0 x $10^{14}$ | 1.24 x $10^6$ |
| X-rays | 3.0 x $10^{13}$ | 1.24 x $10^5$ |
| UV | 1.0 x $10^9$ | 4.1 |
| Visible | 6.0 x $10^8$ | 2.5 |
| IR | 3.0 x $10^6$ | 0.012 |
| Microwaves | 2450 | 0.00001 |
| Radiowaves | 1 | 1.24 x $10^5$ |

*Source:* Zhao, J. and Yan, W., Microwave-assisted inorganic syntheses, in: *Modern Inorganic Synthetic Chemistry*, Xu, R., Pang, W., Huo, Q. (eds.), Elsevier, Amsterdam, the Netherlands, pp. 173–195, Chapter 8, 2011.

### Dielectric Polarization

Electric dipoles in condensed matter can follow the alternating electric field component of the electromagnetic micro-waves, for example, dipolar molecules such as water do rotate with the changing electric field $2.45 \times 10^9$ times per second for the standard 2.45 GHz frequency, and the resistance to such dipole movements generates a considerable amount of heat.[20] The overall dielectric polarizability of condensed matter contains several accumulative components such as electronic polarization of the atomic electron clouds, atomic polarization of the positively charge atomic cores, polarization of the lattice or molecular dipoles, ionic polarization due to displacements of ions, and interfacial polarization at interfaces such as domain or grain boundaries. Since water or other solvents used for hydrothermal synthesis usually exhibit a strongly polar character of the molecules, the mechanism of dielectric polarization is the most important contribution to the heating effect in microwave-assisted hydrothermal synthesis.

### Dielectric losses (loss Currents)

Electrical conduction due to dielectric losses ($\varepsilon''$) can be understood as a movement of free charge carriers in the material as a response to the applied alternating electric field. The dielectric loss heating effect is by heat dissipation due to Joule heating of the loss currents and depends strongly on the electrical resistance of the material. In metals, a large amount of free and mobile electrons are available, and large loss currents occur mainly at the metal surfaces. In fully insulating materials, losses are low and Joule heating does not contribute significantly to the heating process. In semiconducting materials, Joule heating becomes more important as temperature increases, and the effect of heat conduction increases rapidly and eventually becomes as important as dielectric polarization.[48] The power density per unit volume ($P_{V,e}$) of absorbed microwaves in W/m³ by condensed matter due to dielectric losses may be expressed as shown in Equation 1. It should be noted a that $P_{V,e}$ describes the electrical microwave energy absorbed by the medium, which is related to but not exactly the same as the thermal energy created within the medium. The average absorbed power per volume ($P_{V,e}$) due to heat dissipation from the electric field component of the microwave is given by

$$P_{V,e} = \sigma\ E_{rms}^2 = \omega\varepsilon_0\varepsilon'' E_{rms}^2 = 2\frac{\pi}{\lambda}\sqrt{\frac{\varepsilon_0\varepsilon''^2}{\varepsilon'\mu_0\mu'}}\ E_{rms}^2 \quad (1)$$

where
$\sigma$ is the electrical conductivity of the material
$E_{rms}$ is the internal electrical field of the microwave
$\omega$ and $\lambda$ are the angular frequency and wavelength, respectively, of the microwave propagating in the material

Equation 1 can also be found in the literature carrying an additional factor of ½.[49] Equation 1 in the form presented can be easily deduced though from the well-known law of electrical current heat dissipation $P = U \times I$. It should be noted at this point that Equation 1 is different to conventional heating, where heat transfers most efficiently to materials with high thermal conductivity.[50]

The mechanism of electrical current heat dissipation on its own is commonly utilized as the main heating source in spark plasma sintering processes.

### Magnetic Heating

In theory, magnetic materials may be directly heated by the magnetic component of the electromagnetic microwaves coupling to the material's magnetic moments. The magnetic moments may follow the applied alternating magnetic field from the electromagnetic microwave as is the case, for example, in electron paramagnetic resonance spectroscopy. In the case of ferro- or paramagnetic materials, such magnetic alignment would amplify the external applied magnetic field from the microwave. However, such effects have never been clearly quantified to increase microwave heating efficiency, and the overall amount of heat created by this mechanism may be small as compared to the microwave heating of other mechanisms. In the literature, several reports exist though that suggest that certain magnetic materials can be heated more efficiently with the magnetic component of the electromagnetic microwave.[51–53] This indicates that magnetic resonance could play a role under certain circumstances, but exact quantification of this effect for microwave heating is difficult and not well investigated yet.

### Magnetic losses (inDuction or eDDy Currents)

Magnetic losses consist in magnetically induced electrical currents that emerge in condensed matter according to Faraday's law. The currents induced by the magnetic component of the alternating electromagnetic microwaves are often termed Eddy or Foucault currents.[54–57] The magnetic induction heating effect is based on Joule heating due to such Eddy currents, equivalent to dielectric loss heating. The power ($P_{V,m}$) that is absorbed per unit volume is ambiguous, because quite a lot of different expressions can be found in the literature. This may be debited to the fact that Eddy currents tend to occur predominantly on the surface of conducting materials, which is known as the "skin effect." This implies that the microwave penetration depth is significantly lower for conductors as compared to more insulating materials.[58,59] Possible expressions for the penetration depth $d$ in conducting and insulating materials may be given by

$$d_{Conductor} = \frac{\lambda}{\pi}\sqrt{\frac{\varepsilon'}{2\varepsilon''}} \; ; \; d_{Insulator} = \frac{\lambda}{\pi}\frac{\varepsilon'}{\varepsilon''} \quad (2)$$

This leads to two different expressions for the absorbed power per volume ($P_{V,m}$) from the magnetic part of the microwave by more conducting or insulating materials, starting from a general expression for $P_{V,m}$:

$$P_{V,m} = \frac{\sigma\,\omega^2 d^2 \mu_0^2 \mu'^2 H_{rms}^2}{24} \quad (3)$$

$$P_{V,m(conductor)} = \frac{1}{12}\omega\mu_0\mu' H_{rms}^2 = \frac{1}{6}\frac{\pi}{\lambda}\sqrt{\frac{\mu_0\mu'}{\varepsilon_0\varepsilon'}}\,H_{rms}^2 \quad (4)$$

$$P_{V,m(insulator)} = \frac{1}{6}\frac{\varepsilon'}{\varepsilon''}\omega\mu_0\mu' H_{rms}^2 = \frac{1}{3}\frac{\pi}{\lambda}\sqrt{\frac{\mu_0\mu'\varepsilon'}{\varepsilon_0\varepsilon''^2}}\,H_{rms}^2 \quad (5)$$

where $H_{rms}$ is the internal magnetic field of the microwave.[46] In the case of ferromagnetic materials, additional hysteresis losses can occur. Magnetic loss heating is quite similar to the dielectric loss heating mechanism, because both rely on the effect of Joule heating by electrical currents. Equivalent to the dielectric losses, $P_{V,m}$ corresponds to the magnetic microwave energy absorbed by the medium, which may not be exactly the same as the thermal energy created. Magnetic loss heating on its own is commonly used in many industrial and domestic applications such as magnetic induction heated stoves.

### SeconD-OrDer Effects

Many research groups are currently studying a number of anomalies involved with microwave heating that have been broadly termed "nonthermal effects," which include all differences between the microwave and conventional heating methods that cannot be explained by differences in the temperature and the temperature profile.[60] The debate focuses on anomalous effects between the microwave and the surface of the medium that possibly involves the formation of a plasma, and in the case of solids, the solid diffusion may be increased due to such second-order effects.[61,62] It is possible that these phenomena contribute to the great efficiency of microwave synthesis techniques. The effects of dielectric polarization and dielectric/magnetic loss currents are much better understood as compared to possible magnetic resonance and nonthermal effects. Therefore, most of the discussion in the literature focuses on more established mechanisms.

The occurrence of so many different contributions to the microwave heating effect may again be a plausible explanation for its great efficiency, when in fact several heating mechanisms may contribute at the same time. On the other hand, it makes controlled chemical reactions difficult, because it is usually not straight forward to identify to which extent each mechanism contributes to the heating and, therefore, adjustment of the microwave parameters such as power and time are trial-and-error processes. As mentioned earlier, several parameters in Equations 3 and 4 may in fact be temperature dependent, which implies that the coupling efficiency between the microwave and the reactants may change during the heating process.

The wavelength at which industrial and domestic micro-wave ovens operate is standardized to 12.25 cm (2.45 GHz), but other frequencies exist. The choice of 2.45 GHz is motivated by several requirements: First, any interference with telecommunication, wireless networks, or cellular phones has to be avoided. Second, 2.45 GHz is adequate for kitchen microwaves because the associated 12.25 cm wavelength is smaller than the cooking chamber.[22] However, this frequency is not optimized for the heating of water, because liquid water has a higher resonance frequency of 18 GHz. Therefore, in a microwave-assisted hydrothermal synthesis process, the most effective conversion of microwave energy into thermal energy would occur at 18 GHz.[22,63]

In the literature, condensed matter is usually classified into microwave reflecting, absorbing, and transmitting media, based on the penetration depth expressed in Equation 2 and the concomitant energy absorption (Figure 9).

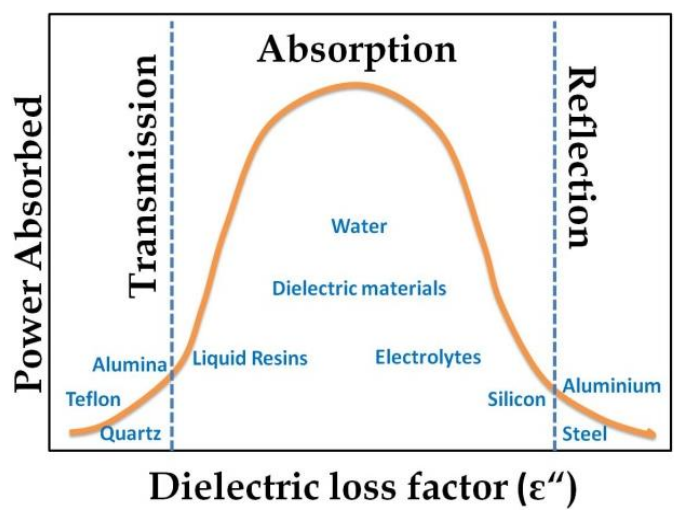

**FIGURE 9** Relationship between dielectric loss $\varepsilon''$ and microwave absorption.

**570**

## CONCLUSIONS

The combination of hydrothermal and microwave synthesis techniques may be a promising approach to significantly reduce the temperature, time, and costs in chemical synthesis. This is interesting not only from industrial aspects but also from a scientific point of view. The reduced heat exposure during synthesis leads to nanosized resultant powder products, where the sizes and shapes of the particles may cover a wide range. This fosters the study of novel nanosized materials with potential for device application.

The synthesis technique of microwave hydrothermal synthesis involves the careful adjustment of several process parameters, such as the microwave parameters power and heating time, as well as the hydrothermal parameters such as temperature, pressure, pH of the solution, and possibly the speed of the magnetic stirrer. All such parameters are difficult to predict for the development of efficient and reproducible microwave hydrothermal synthesis procedures, and in praxis, they need careful adjustment by trial-and-error processes. Therefore, the development of a specific synthesis procedure for a tailored resultant powder product may require a considerable amount of optimization work. On the other hand, the possibility to vary several key process parameters allows changing the synthesis reaction conditions significantly in terms of the reaction kinetics, which in turn gives access to a wide range of different particle sizes and shapes of the same crystalline phase in the resulting product.

Furthermore, the application of microwave hydrothermal synthesis in industrial environments may constitute an important contribution to the development of a *green chemistry* concept in industrial synthesis technology.

## ACKNOWLEDGMENTS

The authors acknowledge funding from the Community of Madrid (Materyener S2009/PPQ-1626) and the Ministry of Science and Innovation (MAT2010-19460). R.S. acknowledges a Ramon y Cajal Fellowship from the MICINN-MINECO in Spain.